\documentclass[12pt,preprint]{aastex}
\usepackage[section] {placeins}
\shorttitle{LiH IN THE ISM}
\shortauthors{Friedel, Kemball, \& Fields}

\begin{document}
\newcommand\pkse{PKS1830-211}
\newcommand\pkso{PKS0201+113}
\newcommand\bo{B0218+357}
\newcommand\Nlih{$1.4\times10^{12}$ cm$^{-2}$}
\title{THE SEARCH FOR EXTRAGALACTIC LITHIUM HYDRIDE}

\author{D. N. Friedel\altaffilmark{1}, Athol Kemball\altaffilmark{1}, \& Brian D. Fields\altaffilmark{1}}

\altaffiltext{1}{Department of Astronomy, 1002 W. Green St., University of
Illinois, Urbana IL 61801\\
email: friedel@astro.illinois.edu}

\begin{abstract}
We have conducted Combined Array for Research in Millimeter-wave Astronomy (CARMA) observations of LiH, in absorption, toward three quasars. These quasars, \bo, \pkse\, and \pkso, have redshifts of $z = 0.685 - 3.387$, and shift the LiH $J=1-0$ transition to the 1 mm and 3mm wavelength bands, where atmospheric absorption is sharply reduced from that predominating near the rest frequency of 443 GHz. We report a 3$\sigma$ detection of LiH toward \bo\ with a column density of \Nlih\ and place an upper limit on the $^6$Li/$^7$Li ratio of $<0.16$. LiH was not detected toward any other source.

\end{abstract}


\section{Introduction}
Big bang nucleosynthesis (BBN) remains our earliest probe of the
universe, and plays a central role in post-WMAP cosmology. Light
element observations and BBN theory measure the cosmic baryon density
($\rho_B \propto \Omega_Bh^2$ ); comparison with the independent WMAP
measure of $\Omega_bh^2$ fundamentally tests the hot big bang
\citep{spergel03,larson}. The result: deuterium observations
spectacularly agree with WMAP, as does helium; in sharp contrast, the
predicted primordial Li is a factor $\ge$ 2 above the best estimates
of the Li abundance in Galactic halo star atmospheres
\citep{cyburt03,coc04,cfo2008}. This discrepancy, between the CMB-predicted primordial Li and
the stellar abundance determinations, is the primordial lithium problem.

The lithium problem and possible solutions have recently been reviewed by \citet{fields}
and \citet{pp}.
A serious possibility is that nonstandard processes occurred during
primordial nucleosynthesis, altering the lithium abundance; dark matter decays can do this \citep{jedamzik06}. 
A signature which would test this
scenario is the production of the $^6$Li isotope at levels
$^6$Li/$^7$Li $\sim 0.05 - 0.10$ \citep{pospelov07,cyburt06}, a ratio
orders of magnitude larger than is found in standard BBN. Indeed,
recent observations of lithium isotopes in halo stars suggest $^6$Li
at these levels \citep{asplund06}. However, these results are highly
controversial due to the difficulty of extracting isotope ratios from
the thermally-broadened 6707 \AA\ Li I line.

The lithium problem may be even worse if there are additional sources
of Li which leave traces in the halo stars where Li is observed. A
very likely source of this kind is Li produced by the interactions of
pre-Galactic or ``structure formation'' cosmic rays (SFCRs) which
should be accelerated in cosmic shocks which accompany the infall of
baryonic gas onto protogalaxies \citep[e.g.][]{miniati00}. The
resulting non-primordial $^6$Li and $^7$Li production can be
significant compared to their primordial levels \citep{fields05} and
would need to be subtracted from halo star abundances to determine the
true primordial levels. This both exacerbates the lithium problem but
also opens the possibility of using Li abundances in different cosmic
environments to probe both the early universe and cosmic-ray
acceleration in structure formation.

Lithium also plays a role in the primordial chemistry leading to the
formation of the first stars.
Namely, the lithium hydride (LiH) molecule
\citep{bg97,gp98}
contributes to the cooling of the dust-free primordial gas.
Current calculations suggest however that the impact of LiH is not dominant due to the small
primordial Li/H abundance, but
the chemistry of primordial LiH is complex and a subject of ongoing interest
\citep{bovino}.

Lithium thus takes a uniquely important role in the cosmology of the early universe. Unfortunately, lithium is difficult to measure due to its tiny abundance: in the solar system, Li/H $\sim$ 10$^{-9}$, while in metal-poor halo stars, Li/H $\sim$ (1-2)$\times$10$^{-10}$. To date, optical spectroscopy of Galactic stellar atmospheres has been the only probe of Li in the early Galaxy, and stubborn systematic uncertainties could account for the Li problem. Any alternative means of observing Li in any source would be extraordinarily useful for many reasons, not least of which would be the different systematics. Furthermore, to date there are no stellar observations of extragalactic Li, and so any extragalactic information would shed qualitatively new light onto this problem.

Since the atomic lines of Li have been so challenging to observe, we have chosen to search for the simplest Li containing molecule, lithium hydride (LiH). LiH has been searched for previously by \citet{combes98} toward \bo, where the authors report a tentative detection in a system that is 5 km/s offset from the systemic velocity but within the CO velocity profile.

\section{Sources}
The LiH molecular isotopomers have a {\it v}=0, {\it J}=0-1 rotational transition at 443.95293 GHz ($^7$LiH) and 453.16028 GHz ($^6$LiH) \citep{bellini94,plummer84}. This region of the mm-wave spectrum is highly opaque through the atmosphere. However these lines are accessible in the 1 \& 3 mm bands at the Combined Array for Research in Millimeter-wave Astronomy (CARMA) for extragalactic sources in redshift ranges of {\it z}=0.64-1.1 and 2.85-4.2, respectively.

Our strategy for detecting LiH in these redshift ranges is to search for interstellar LiH absorption toward lensed quasars with high millimeter continuum flux density. The absorption originates from molecular clouds in the lensing galaxy along the line of sight. These galaxies should preferentially exhibit high optical depth in other atomic and/or molecular transitions, including H and CO. There are only a few candidates known to date. The three best candidate lenses which meet these criteria are B0218+357, PKS1830-211, and PKS0201+113. Table~\ref{tab:mols} lists the source redshift, and red-shifted rest frequencies of our search lines ($^7$LiH, $^6$LiH, and $^{13}$CO). 

\begin{deluxetable}{llrrrcrcc}
\tablecolumns{5}
\tablewidth{0pt}
\tablecaption{Frequencies of Observed Lines\tablenotemark{a}}
\tablehead{\colhead{Source} & \colhead{{\it z}} & \colhead{$^7$LiH\tablenotemark{b}} &
\colhead{$^6$LiH\tablenotemark{c}} & \colhead{$^{13}$CO\tablenotemark{d}} & \colhead{Ref}}
\startdata
B0218+357   & 0.68466 & 263.526648 & 268.992129 & 261.634494 & 1\\
PKS1830-211 & 0.88582 & 235.416319 & 240.298798 & 233.726001 & 2\\
PKS0201+113 & 3.387144 & 101.194035 & 103.292775 & 100.467449 & 3\\
\enddata
\tablenotetext{a}{All frequency units are GHz.}
\tablenotetext{b}{The unshifted rest frequency of the $J=0-1$ $^7$LiH transition is 443.95280 GHz}
\tablenotetext{c}{The unshifted rest frequency of the $J=0-1$ $^6$LiH transition is 453.16028 GHz}
\tablenotetext{d}{The unshifted rest frequency of the $J=3-4$ $^{13}$CO transition is 440.7651668 GHz}
\tablerefs{(1) \citet{combes98}; (2) \citet{muller08};  (3) \citet{kanekar07}}
\label{tab:mols}
\end{deluxetable}

\subsection{B0218+357}
\bo\ is a gravitational lens with an Einstein ring. The source is a BL Lac object at a redshift {\it z}$\sim$0.94, and appears, at arcsecond resolution, as two distinct point sources in the ring, with a separation of 335 mas, and a weak jet called the ``hot spot''\citep{patnaik93,odea92,wiklind95,biggs03,cohen03}. The point sources, A \& B, have a flat spectrum, while the Einstein ring has a steep spectrum. They are highly variable on a timescale of a few days and each is separated into at least two subcomponents separated by a few mas \citep{biggs01,patnaik95}. The A source is brighter by a factor of $\sim$3 in the radio and both have jets \citep{wiklind95,biggs01}. The lensing source is a spiral galaxy at a redshift of {\it z}$\sim$0.685 \citep{patnaik95, wiklind95}. Atomic and molecular lines including the 21 cm H line \citep{carilli93}, H$_2$CO \citep{menten96}, H$_2$O, CS \citep{combes97}, CO, HCO$^+$, HCN \citep{wiklind95}, NH$_3$ \citep{henkel05}, OH \citep{kanekar03}, H$_2$CO anti inversion \citep{zeiger10}, and LiH (tentative) \citep{combes98} have been detected toward the A component. The line center of the LiH tentative detection is off by $\sim$5 km/s compared to other detected lines, but within the CO velocity profile \citep{combes98}.

\subsection{PKS1830-211}
\pkse\ was one of the first quasars that was found to be lensed. The source blazar is at a redshift of {\it z}$\sim$2.5 and appears as two bright images in a faint Einstein ring at sub-arcsecond resolution \citep{lidman99,subra90}. The two images are NW and SE of the lens center and are separated by $\sim$1\arcsec; the NW component is the brighter of the two \citep{jin99}. The images vary in brightness, separation, and size due to a helical jet emanating from the core \citep{jin03,nair05}. These variations cause notable changes in absorption line features in the lensing galaxy \citep{muller08}. The lens galaxy is a face on spiral at a redshift of {\it z}$\sim$0.886 \citep{winn02,wiklind98}. One of the spiral arms lies along the line of sight to the SW image and is the primary source of molecular and atomic absorption lines, although a small percentage also originates from the NW source \citep{winn02}. Numerous molecular lines have been detected toward \pkse\ including CO, HCN, HCO$^+$ \citep{wiklind95}, CS, H$_2$O \citep{combes97}, N$_2$H$^+$ \citep{wiklind98}, C$_2$H, HC$_3$N, C$_3$H$_2$ \citep{menten99}, NH$_3$ \citep{henkel05}, and H$_2$CO \citep{menten96}.

\subsection{PKS0201+113}
\pkso\ is a radio loud quasar at a redshift of {\it z}$\sim$3.61 \citep{condon77,white93}. VLBI observations indicate that its linear size is small $<$40 pc and has a smaller secondary component 2\arcsec south \citep{hodges84,morabito86,stanghellini91}. The secondary source emits on 1\% of the total system flux (at 22 GHz). \pkso\ shows damped Lyman $\alpha$ along the line of sight, several atomic lines, and 21 cm absorption at {\it z}$\sim$3.387 \citep{white93,oya98,kanekar07}.

\section{Observations}
All observations were conducted with the Combined Array for Research in Millimeter-wave Astronomy (CARMA). In all cases the source is unresolved in our beams, and thus appears as a point source. \pkso\ was observed in 2008 February when CARMA was in its B configuration giving a typical synthesized beam of 0.88$\times$0.70\arcsec. The $\it u-v$ coverage of the observations gives projected baselines of 25.2-319.6 k$\lambda$ (74.6-946.9 m). All three correlator bands were set up with 62 MHz bandwidths and positioned with overlapping edge channels. This gives full frequency coverage across the {\it J} = 0-1 $^7$LiH line with a velocity resolution of $\sim$2.9 km/s. Uranus was used as a flux density calibrator, the internal noise source was used to correct the passbands of each window, and 0238+166 was used to calibrate the complex antenna-based gains. The source was then self-calibrated to refine the solution. The data were calibrated and imaged using the MIRIAD software package \citep{sault95}. We obtained about ten hours of good data on this source.

\pkse\ was observed in 2008 March and April, when CARMA was in its D configuration giving a typical synthesized beam of 3.6$\times$2.1\arcsec. The {\it u-v} coverage of the observations gives projected baselines of 3.79-95.03 k$\lambda$ (4.8-121.1 m). All three correlator bands were set up with 62 MHz bandwidths, giving a resolution of $\sim$1.2 km/s. Two of the bands were configured to overlap at the edges to cover the {\it J} = 0-1$^7$LiH line in the lower sideband and {\it J} = 0-1 $^6$LiH in the upper sideband. The remaining band was configured to observe the {\it J} = 3-4 $^{13}$CO line in the lower sideband, this choice was driven by the frequency configuration. Neptune and MWC349 were used as flux density calibrators, 3C273 and 3C454.3 were used to correct the passbands of each window, and 1733-130 and 1911-201 were used to calibrate the antenna based gains. The source was then self-calibrated to refine the solution. We obtained only 4.5 hours of good data due to poor weather. 

\bo\ was observed in 2008 August and September, when CARMA was also in its D configuration, giving a typical synthesized beam of 1.7$\times$1.5\arcsec. The $u-v$ coverage of the observations gave projected baselines of 6.7-129.2 k$\lambda$ (7.7-147.1 m). Two of the correlator bands were set up in 62 MHz mode, giving $\sim$1.1 km/s resolution, and the third was in 31 MHz mode, giving $\sim$0.55 km/s resolution. Band 1 in the lower sideband was centered on the {\it J} = 3-4 $^{13}$CO line. Bands 2 and 3 in the lower sideband were set to observe the {\it J} = 0-1 $^7$LiH transition. Band 1 in the upper sideband was set to observe the {\it J} = 0-1 $^6$LiH transition. The upper sideband of the other two bands were for continuum. Uranus was used as the flux density calibrator, 3C84, 3C454.3, and the internal noise source were used to correct the passbands of each window. The data were self-calibrated for the antenna based gains. We only obtained about seven hours of good data on this source due to poor weather.

\section{RESULTS \& DISCUSSION}
\subsection{\bo}
\begin{figure}[ht!]
\includegraphics[angle=270,scale=0.8]{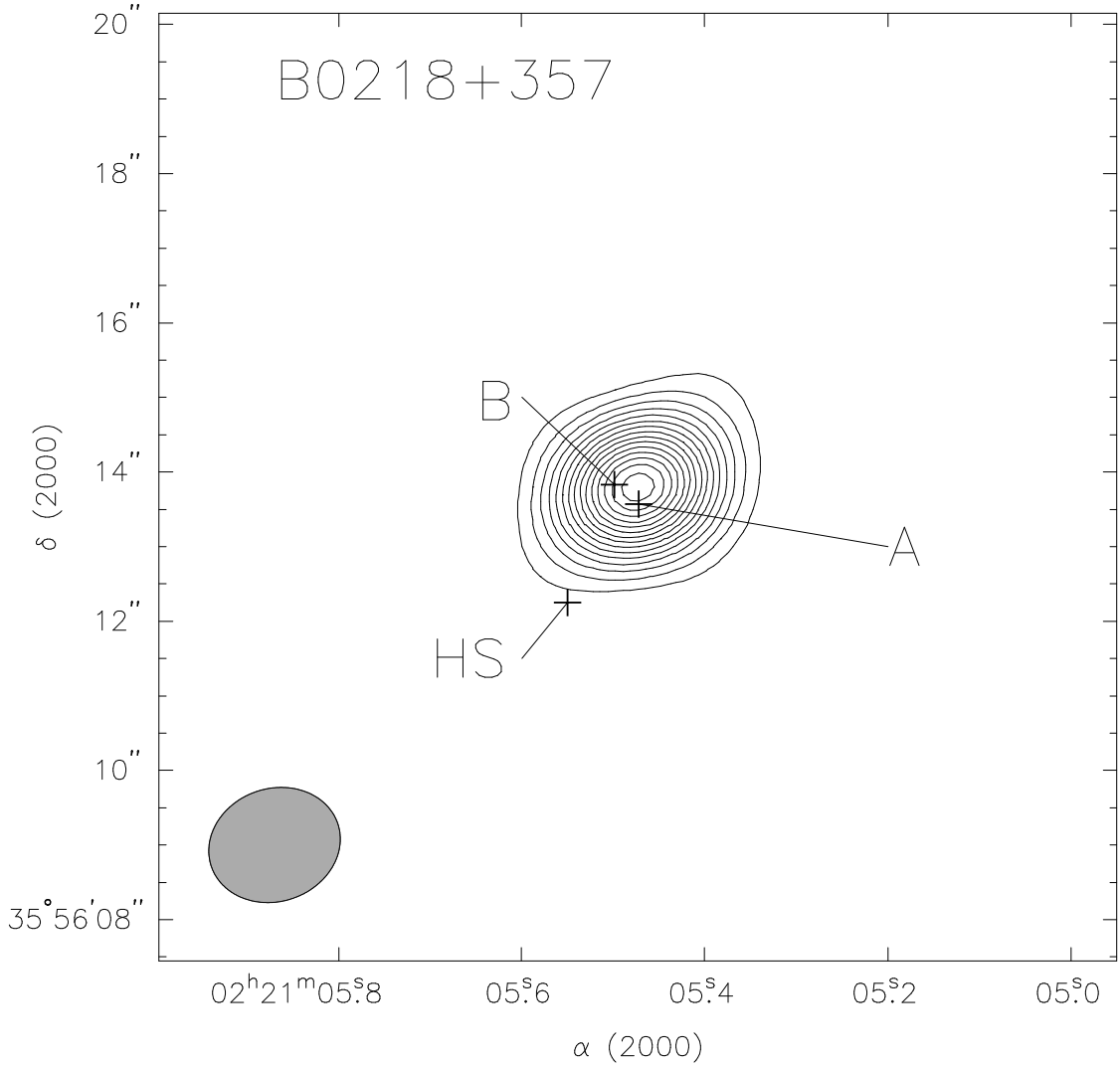}
\caption{\bo\ continuum map. The contours are $\pm$12$\sigma$, $\pm$24$\sigma$, $\pm$36$\sigma$, ..., $\sigma$ = 4.5 mJy/beam. The synthesized beam is plotted in the lower left corner and the known continuum sources (A and B) and the 'hot spot' (HS) are labeled.\label{fig:bo-contin}}
\end{figure}

\begin{figure}[ht!]
\includegraphics[angle=270,scale=0.8]{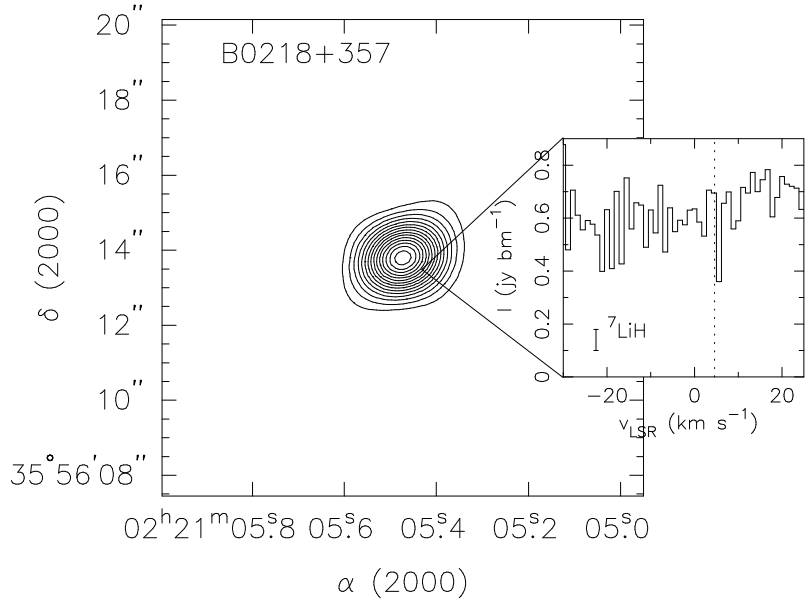}
\caption{LiH detection toward \bo. The noise level of the spectrum is denoted by the error bar in the lower left corner. The dotted line denotes the expected rest velocity of the source.\label{fig:bo-lih}}
\end{figure}

Figure~\ref{fig:bo-contin} shows our continuum map of \bo. The two point sources (A and B) and the ''hot spot'' are labeled and the synthesized beam is in the lower left corner. The contour levels are $\pm$12$\sigma$, $\pm$24$\sigma$, $\pm$36$\sigma$, ..., $\sigma$ = 4.5 mJy/beam and the peak flux is 885 mJy/beam. Note that we do not detect the ''hot spot'' (even at our 3 $\sigma$ cutoff). This peak flux value is within the uncertainties of the expected value at our rest frequency from Figure~2 of \citep{combes97}. Figure~\ref{fig:bo-lih} shows our 3$\sigma$ detection of $^7$LiH. The noise level of the spectrum is denoted by the error bar and the dashed line denotes the expected rest velocity of the line. This line was detected toward the A component, where all other line detection have been made to date and is at the expected rest velocity. For absorption features the total beam averaged column density can be calculated from \citep{ll93,greaves96b}

\begin{equation}
\langle N_T\rangle=\frac{8.0Q_re^{E_l/T_{ex}}\int\tau dv}{S\mu^2(1-e^{-h\nu/kT_{ex}})}\times10^{12}~{\rm cm}^{-2},
\label{eqn:ntabs}
\end{equation}
where $Q_r$ is the rotational partition function, $E_l$ is the lower state energy, $S\mu^2$ is the product of the line strength and dipole moment, and the opacity, $\tau$, is defined below

\begin{equation}
T_{MB}=(1-e^{-\tau})(J(T_{ex})-J(CB)-T_C)
\end{equation}
giving
\begin{equation}
\tau=-{\rm ln}\left[1-\frac{T_{MB}}{J(T_{ex})-J(CB)-T_C}\right],
\end{equation}
where $T_{MB}$ is the main beam brightness temperature of the line, $J(T)$ is defined as \citep{rw00}
\begin{equation}
J(T)=\frac{h\nu}{k}\frac{1}{e^{h\nu/kT}-1}~{\rm K},
\end{equation}
$J(CB)$ is the cosmic microwave background ($2.73(1+z)$), and $T_C$ is the continuum brightness temperature. For observations not done in a temperature scale one can convert from Jy beam$^{-1}$ to K using \citep{rw00}
\begin{equation}
T=\frac{1.22I_0}{B\theta_a\theta_b\nu^2}~{\rm K},
\label{eqn:i2t}
\end{equation}
where $I_0$ the is peak intensity of the line in Jy beam$^{-1}$ and $B$ is the beam filling factor (incorporating both the continuum and absorption source factors). In all equations the excitation temperature $T_{ex}$ is assumed to be 10 K.

From these equations we get an opacity ($\tau$) of 1.7 and a total column density of \Nlih\ for $^7$LiH. This value is similar to that of \citep{combes97}. Calculated H$_2$ column densities for this source range from $5\times10^{21}$ - $5\times10^{23}$ cm$^{-2}$ \citep{menten96,combes95}. From these values we calculate a LiH/H$_2$ ratio of $2.9\times10^{-10}$ - $2.9\times10^{-12}$. If we assume that the gas is cold and dark, as did \citet{combes98}, we have f(H$_2$) = 0.5. The Li/H ratio for a source with the redshift of \bo\ can be estimated to be $\sim10^{-9}$ \citep{combes98}. From these values we calculate a LiH/Li ratio of 0.287 - 0.003.

\begin{figure}[ht!]
\includegraphics[angle=270,scale=0.8]{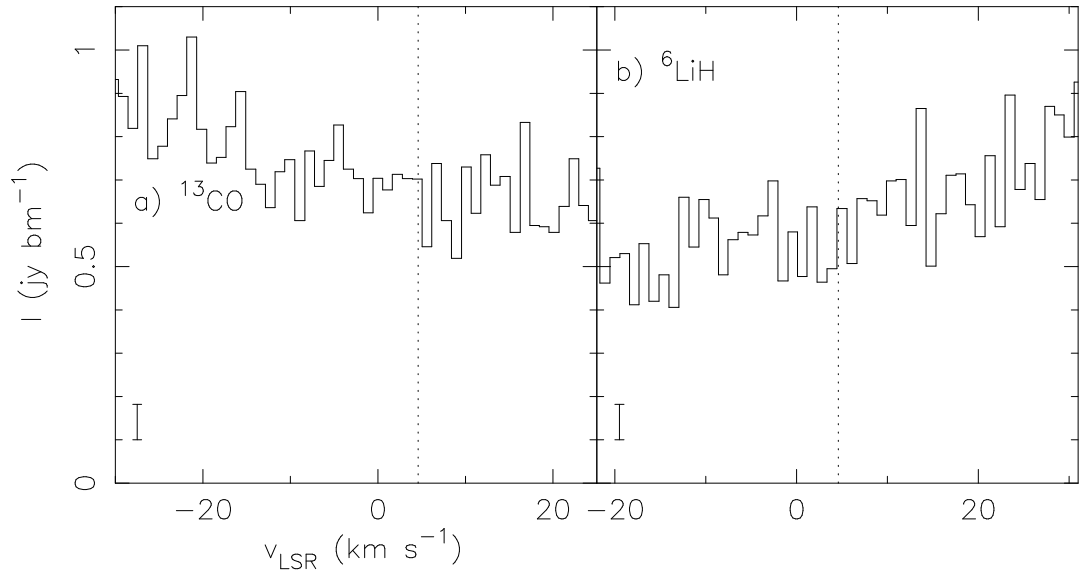}
\caption{Non detections toward \bo. The dashed line denotes the expected rest velocity of the lines and the error bar gives the rms noise level for each spectra. a) $^{13}$CO non detection b) $^6$LiH non detection.\label{fig:bo-non}}
\end{figure}

Figure~\ref{fig:bo-non} shows our non-detections toward \bo. a) shows $^{13}$CO and b) shows $^6$LiH. Using equation~\ref{eqn:ntabs} we can set 1$\sigma$ upper limits of $4.8\times10^{15}$ cm$^{-2}$ and $2.3\times10^{11}$ cm$^{-2}$, respectively. From this we can calculate a $^6$Li/$^7$Li upper limit of $<$0.16. The $J=2-1$ $^{13}$CO transition was detected by \citep{combes95}, thus given our non-detection we can set limits on the temperature. Given our 1 $\sigma$ upper limit we can only set a temperature upper limit of 75 K.

\subsection{\pkse}
\begin{figure}[ht!]
\includegraphics[angle=270,scale=0.8]{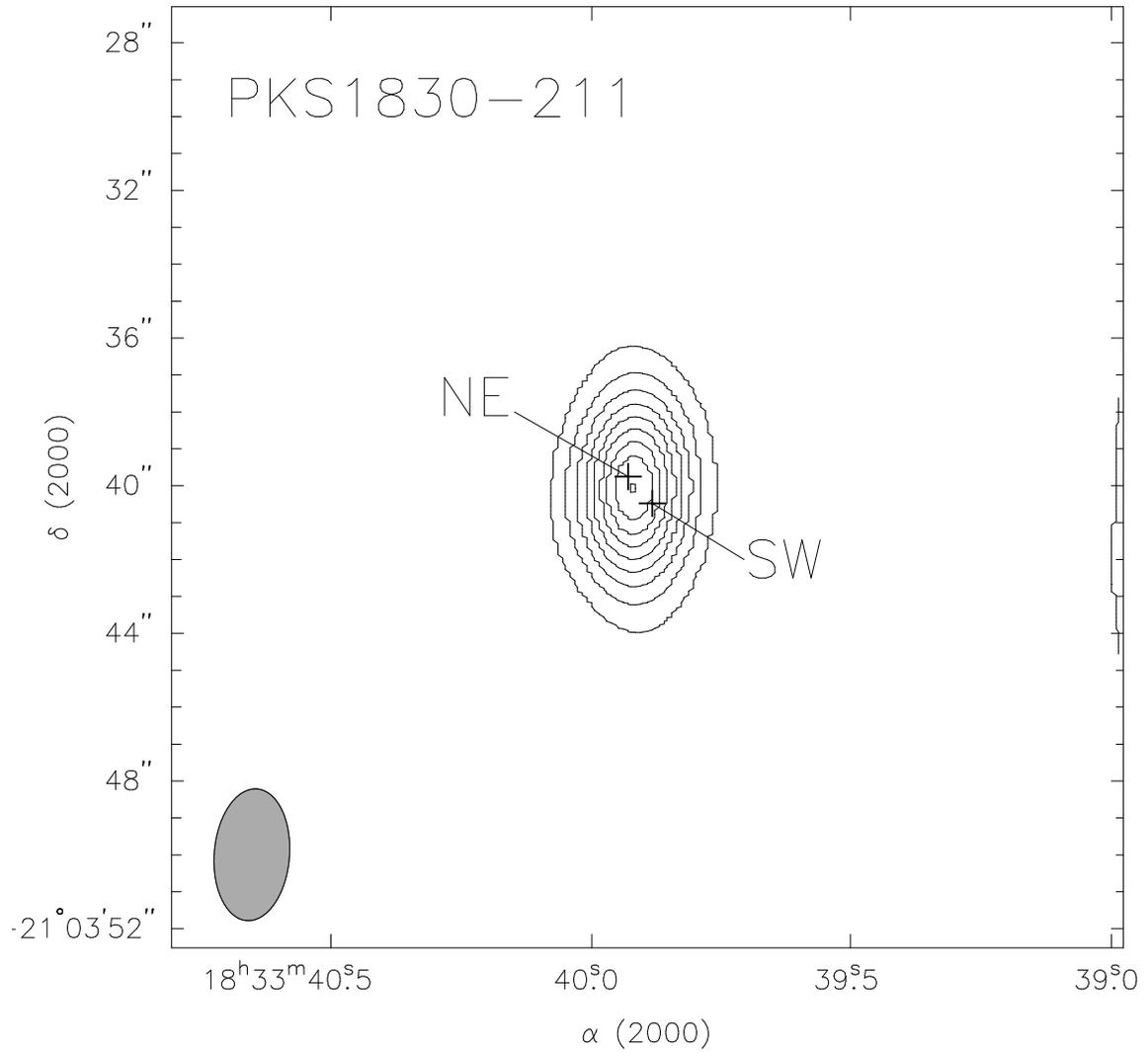}
\caption{\pkse\ continuum map. The contours are $\pm$12$\sigma$, $\pm$24$\sigma$, $\pm$36$\sigma$, ..., $\sigma$ = 5.6 mJy/beam. The two point sources (NW and SE) are labeled. The synthesized beam is in the lower left corner.\label{fig:pkse-contin}}
\end{figure}

\begin{figure}[ht!]
\includegraphics[angle=270,scale=0.8]{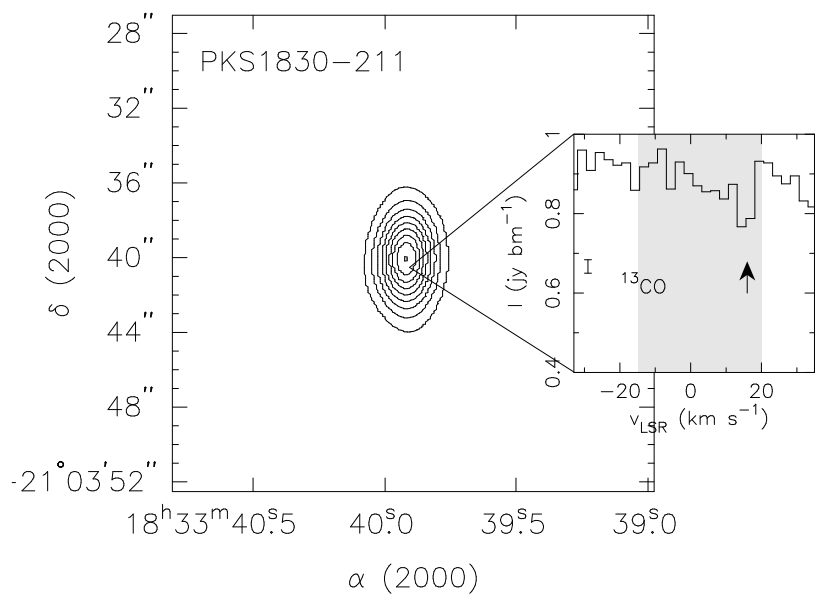}
\caption{$^{13}$CO toward \pkse. The error bar denotes the 1$\sigma$ noise level of the spectra and the grey region indicates the range of expected rest velocities from \citet{muller06,menten08}.\label{fig:pkse-co}}
\end{figure}

Figure~\ref{fig:pkse-contin} shows our continuum map of \pkse. The two point sources (NE and SW) are labeled. The synthesized beam is in the lower left corner and the contour levels are $\pm$12$\sigma$, $\pm$24$\sigma$, $\pm$36$\sigma$, ..., $\sigma$ = 9.0 mJy/beam and the peak flux is 953 mJy/beam. Figure~\ref{fig:pkse-co} shows our 3.4$\sigma$ detection of $^{13}$CO toward the SW component of \pkse. The error bar denotes the rms noise level of the spectrum. The grey area indicates the velocities over which absorption peaks have been detected previously \citep[e.g.][]{muller06,menten08}. The arrow indicates the line center. In order to increase the S/N of our spectra we inverted the data by averaging adjacent together, decreasing the velocity resolution to 2.5 km/s.

From equation~\ref{eqn:ntabs} we calculate a column density of $2.8\times10^{16}$ cm$^{-2}$ for $^{13}$CO. Comparing this value with the $^{12}$CO value from \citet{wiklind98} gives a $^{12}$CO/$^{13}$CO ratio of $\sim$72. While this value is a bit high compared to most Galactic values it is not unreasonable, since $^{13}$C is formed primarily in low and medium mass stars during their red giant phase, and there are fewer of these at higher {\it z}, one could reasonably expect a higher ratio toward \pkse\ \citep[][and references therein]{wr94}.

\begin{figure}[ht!]
\includegraphics[angle=270,scale=0.8]{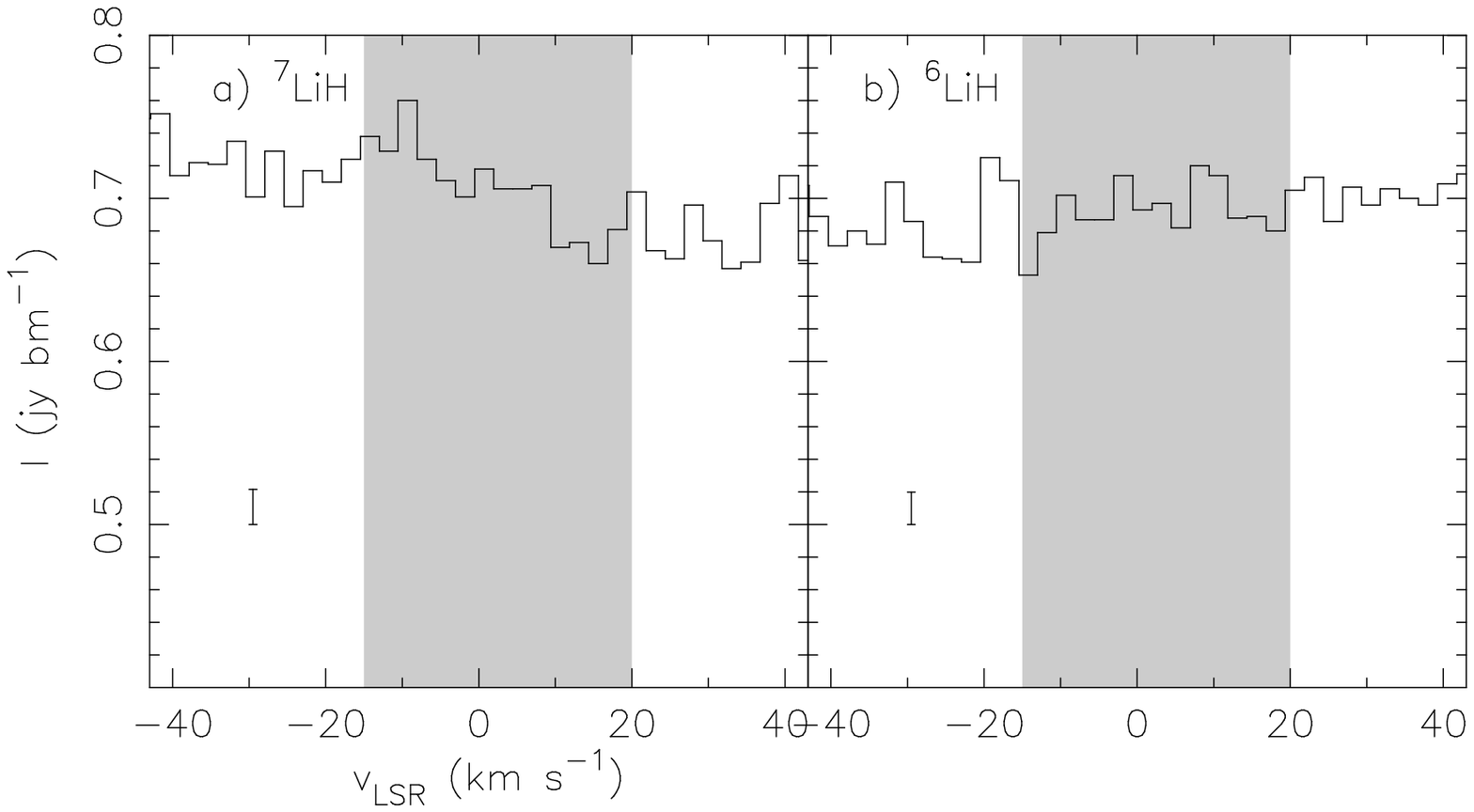}
\caption{Non detections toward \pkse. The grey region indicates the range of expected rest velocities from \citet{muller06,menten08} and the error bar gives the rms noise level for each spectra. a) $^{7}$LiH non detection b) $^6$LiH non detection.\label{fig:pkse-non}}
\end{figure}

Figure~\ref{fig:pkse-non} shows our non-detections toward \pkse. a) shows $^{7}$LiH and b) shows $^6$LiH. Using equation~\ref{eqn:ntabs} we can set 1$\sigma$ upper limits of $3.4\times10^{10}$ cm$^{-2}$ and $3.7\times10^{10}$ cm$^{-2}$, respectively.

\subsection{\pkso}
\begin{figure}[ht!]
\includegraphics[angle=270,scale=0.8]{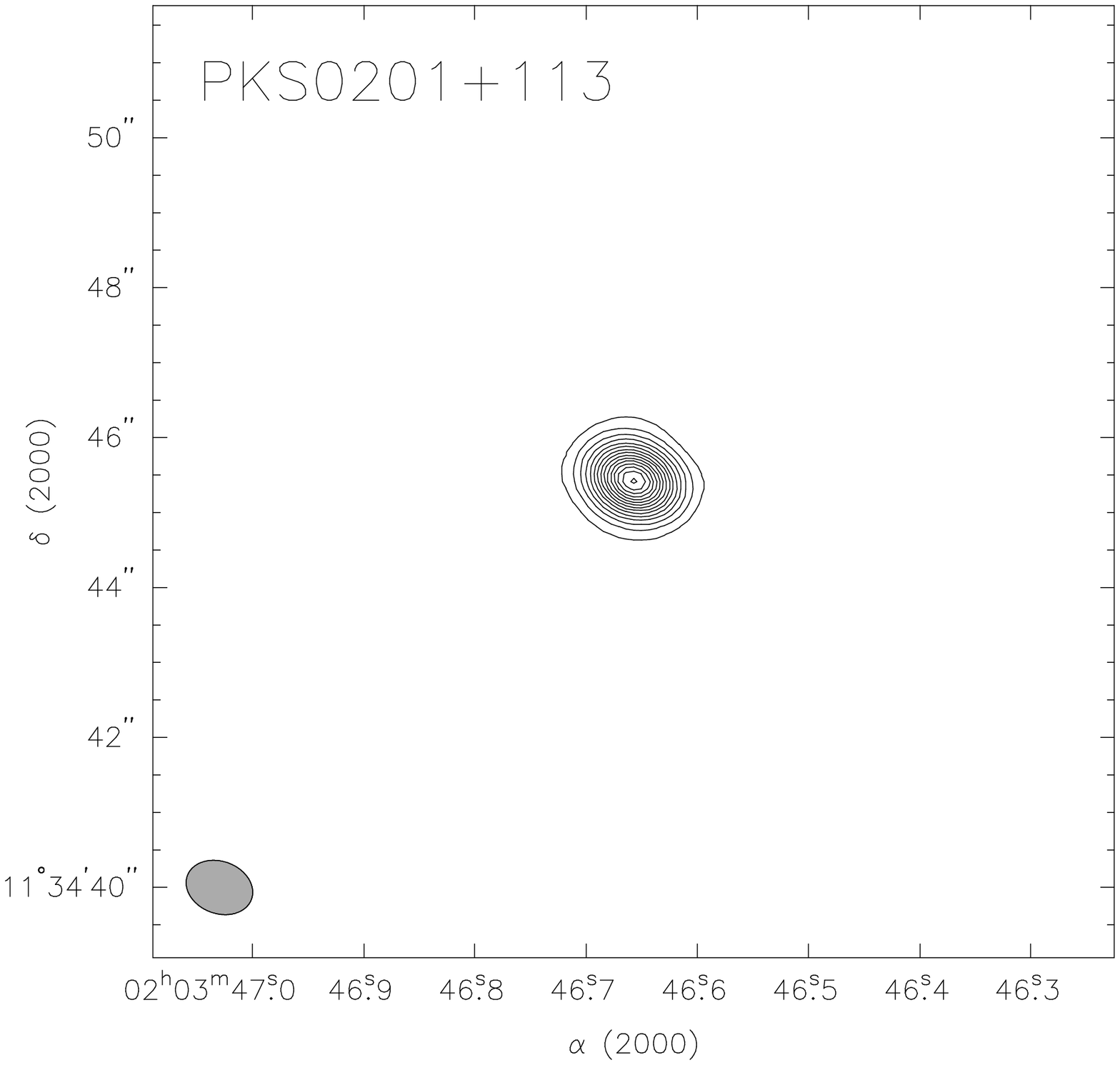}
\caption{\pkso\ continuum map. The contours are $\pm$12$\sigma$, $\pm$24$\sigma$, $\pm$36$\sigma$, ..., $\sigma$ = 926 $\mu$Jy/beam. The synthesized beam is plotted in the lower left corner.\label{fig:pkso-contin}}
\end{figure}

Figure~\ref{fig:pkso-contin} shows our continuum map of \pkso. The synthesized beam is plotted in the lower left corner and the contour levels are $\pm$12$\sigma$, $\pm$24$\sigma$, $\pm$36$\sigma$, ..., $\sigma$ = 926 $\mu$Jy/beam and the peak flux is 169 mJy/beam.

\begin{figure}[ht!]
\includegraphics[angle=270,scale=0.8]{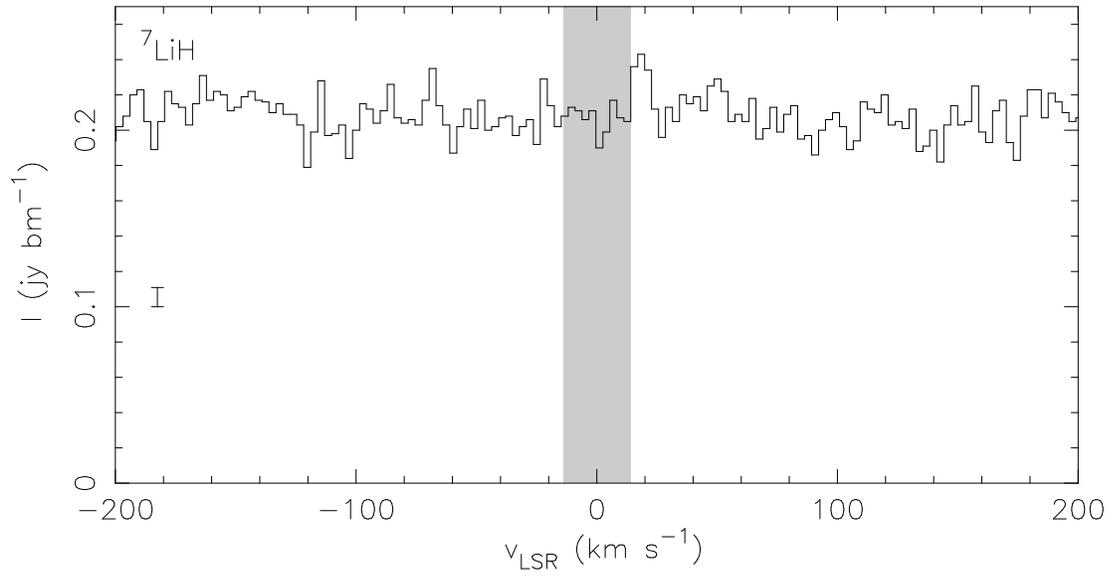}
\caption{Non detection of $^7$LiH toward \pkso. The grey region indicates the range of expected rest velocities from \citet{kanekar07} and the error bar denotes the 1$\sigma$ rms noise level.\label{fig:pkso-lih}}
\end{figure}

We did not detect $^7$LiH toward this source. Figure~\ref{fig:pkso-lih} shows the spectrum of the non-detection. The error bar denotes the 1$\sigma$ rms noise level. From equation~\ref{eqn:ntabs} we calculate an upper limit to the column density of $1.0\times10^{10}$ cm$^{-2}$.

\section{Implications and Conclusions}
We have presented CARMA observations of the $J=0-1$ transitions of $^7$LiH and $^6$LiH, and the $J=3-4$ transition of $^{13}$CO toward three distant quasars, \bo, \pkse, and \pkso. While these lines were not detected in most of the sources, due primarily to the lack of sensitivity, some were detected. The $^7$LiH line was detected toward only one source, \bo, at a $\sim$3-$\sigma$ level. From these observations we calculate a beam averaged column density of \Nlih. We also searched for the $J=0-1$ transition of $^6$LiH toward \bo, but did not detect it. We were able to set an upper limit on the $^6$Li/$^7$Li ratio of $<$0.28 for this source. Unfortunately, this upper limit does not indicate which model for BBN is correct. Further observations to detect the $^6$LiH transition are necessary to resolve this. 

Ultimately one wants to know the primordial Li abundance. Observing and detecting LiH is just the first step in this process. The next step is determining the Li column density from the LiH. Doing this requires detailed knowledge of the formation and destruction processes of LiH, both in the gas phase (for both high and low $z$ objects) and on grain surfaces (for low $z$ objects). To date this knowledge is notably incomplete, especially over the range in $z$ in which we can observe LiH. Additionally the H column density needs to be determined to good accuracy. Recent determinations of the H column density in many of these sources is based on CO/H ratios, using CO lines that may be saturated, and assuming that the CO/H ratio is constant for all $z$, which is not likely to be the case. Thus, while LiH offers a new probe of extragalactic lithium, much work is still needed before LiH yields the primordial abundance.

\acknowledgements
We thank an anonymous referee for many helpful comments which improved this manuscript. This work was partially funded by NSF grant AST-0540459 and the University of Illinois. Support for CARMA construction was derived from the states of Illinois, California, and Maryland, the Gordon and Betty Moore Foundation, the Kenneth T. and Eileen L. Norris Foundation, the Associates of the California Institute of Technology, and the National Science Foundation. Ongoing CARMA development and operations are supported by the National Science Foundation under a cooperative agreement, and by the CARMA partner universities.

\end{document}